\documentclass[preprint,proceedings]{rmaa}

\suppressfulladdresses 

\usepackage{paralist}

\usepackage{psfrag,color}

\usepackage{graphics}   

\SetYear{2006}
\SetConfTitle{Massive Stars: Fundamental Parameters and Circumstellar Interactions}

\title{Towards using optical/NIR photometry to measure the temperature of O stars}

\author{
  J. Ma\'{\i}z Apell\'{a}niz\altaffilmark{1}  
                 and A. Sota\altaffilmark{2,3}}

\altaffiltext{1}{Instituto de Astrof\'{\i}sica de Andaluc\'{\i}a-CSIC.}
\altaffiltext{2}{Space Telescope Science Institute.}
\altaffiltext{3}{Universidad Aut\'onoma de Madrid.}

\shortauthor{Ma\'{\i}z Apell\'{a}niz \& Sota}
\shorttitle{Photometry and the temperatures of O stars}

\suppressfulladdresses

\listofauthors{J. Ma\'{\i}z Apell\'{a}niz and A. Sota}
\indexauthor{Ma\'{\i}z Apell\'{a}niz, J.}
\indexauthor{Sota, A.}

\abstract{It has been traditionally stated that it is not possible to use optical/NIR photometry to measure the
temperatures of O stars. In this contribution we describe the steps required to overcome the hurdles that have
prevented this from happening in the past and we present our preliminary results for the low-extinction case.}
\resumen{Tradicionalmente se ha dicho que no es posible usar fotometr\'{\i}a visible/IR para medir la temperatura
de las estrellas O. En esta contribuci\'on describimos los pasos necesarios para vencer los obst\'aculos
existentes hasta la fecha y presentamos nuestros resultados preliminares para el caso de extinci\'on baja.}

\addkeyword{stars: atmospheres}
\addkeyword{stars: early-type}
\addkeyword{stars: fundamental parameters}

\begin{document}

\maketitle 

\section{Background}

	Hummer et al. (1988) declared two decades ago that ``for hot stars ($T_{\rm eff} > 30\,000$ K), methods
based on the integrated continuum flux are completely unreliable discriminators of the effective temperature''.
This statement has been subsequently reformulated to express that optical/NIR broad- or medium-band
photometry cannot be used to determine the intrinsic properties of O stars. However, the data used by Hummer et
al. (1988), as the authors themselves recognize, had observational uncertainties of 0.02 magnitudes in $E(B-V)$
and their flux calibration was only accurate to within 10\% in the UV and 3\% in the optical/NIR. Therefore,
it is a legitimate question to ask whether optical/NIR can indeed be used to measure O-star temperatures if the
precision and accuracy of our data are better than those amounts. In this contribution we present the steps
required to reduce the uncertainties to a 1\% level.

\section{The intrinsic colors of O stars}

\subsection{Filter properties and zero-point determination}

	The first required step is to calibrate the photometry by: (a) accurately measuring the
reference SED (e.g. Vega), (b) determining the zero point that fine-tunes the correspondence between the observed 
and the reference SEDs, and (c) correctly defining the total (atmosphere + telescope + filter + detector) 
sensitivity curve of the system. (a) and (b) are needed in order to avoid systematic errors between different 
filters while (c) is needed to avoid ``color terms'' within a given filter. Some
recent papers have managed to produce significant improvements in all of the above. Bohlin (2006) has used a
combination of STIS spectrophotometry and Kurucz models to obtain a new SED for Vega that eliminates previous 
discrepancies between the results of different groups. Several other papers (Cohen et al. 2003, Ma\'{i}z
Apell\'aniz 2005, 2006a, 2006b, Holberg \& Bergeron 2006) have analyzed different filter systems (Johnson,
Str\"omgren, 2MASS\ldots) to test the validity of the published sensitivity curves and to calculate the zero
points. Interestingly, independent results for the same filter agree to within 1\%, with the only outstanding
exception of 2MASS $K_s$ (Ma\'{\i}z Apell\'aniz 2006b).

\subsection{Atmosphere models}

	The second required step is to be able to use atmosphere models as inputs for the calculation of the
intrinsic colors of O stars. The last five years have seen a significant increase in the detail of such models.
The two most significant additions have been the inclusion of line blanketing by heavy elements and wind
effects, the former being the one that has the largest effect in the optical and the latter the dominant one
in the $K$ band and longer wavelengths (when mass-loss rates are high, see Martins \& Plez 2006). 
There are now grids for early B 
(Lanz \& Hubeny 2006) and O (Lanz \& Hubeny 2003, Martins et al. 2005) stars that cover the 
$T_{\rm eff}-\log g-Z$ space needed to derive synthetic broad-band colors for most hot stars. Furthermore, the
synthetic colors from different models agree to within $\sim 1$\%, lending credence to their reliability and making
them appropriate to be used as inputs for temperature determinations of early-type stars.

\section{Temperature determination techniques: CHORIZOS}

	The traditional method for calculating $T_{\rm eff}$ from photometry has been through color-color diagrams 
and $Q$-parameter determinations. That strategy has several problems: 
[a] the slope of an extinction trajectory in a
color-color plot depends on the initial SED; [b] the slope is not constant, either, as a function of the amount of
extinction; and [c] it only allows the simultaneous use of information from two colors, hence being especially
sensitive to systematic errors in the calibration and to deviations from the expected unextincted SEDs. All of the
above can easily introduce considerable systematic errors in the determination of stellar temperatures but those
problems are minimized when one uses a Bayesian code like CHORIZOS (Ma\'{\i}z Apell\'aniz 2004), which can process
information from many filters simultaneously without using the constant-slope $Q$ approximation. This constitutes
the third required step towards determining $T_{\rm eff}$ for O stars from photometry.


\begin{figure}
\centerline{\includegraphics[width=\linewidth]{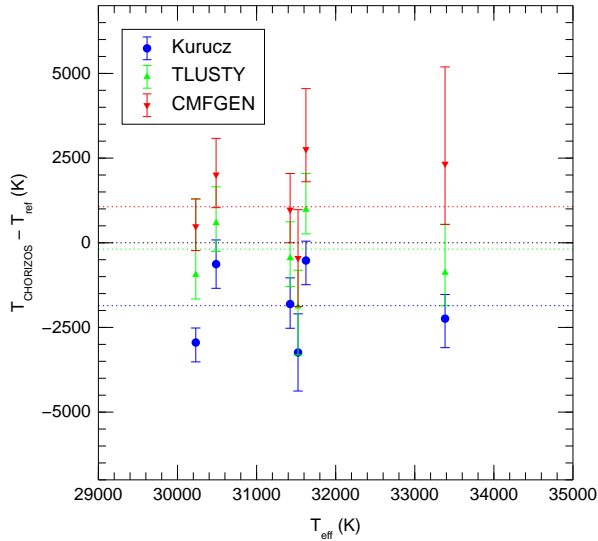}}
\caption{Difference between photometric and spectroscopic temperature determinations for six low-extinction O stars
using different SED families as inputs.}
\end{figure}

	Taking into account all of the previous considerations and as a first test of the use of CHORIZOS to 
determine O-star temperatures, we are working on an analysis of 
optical/NIR photometry of low-extinction ($E(B-V) <$ 0.15) O stars with
CHORIZOS. Our preliminary results for six late-type O stars using Str\"omgren and 2MASS photometry fitting 
temperature, extinction, and extinction law simultaneously are shown in 
Fig.~1. Using either TLUSTY or CMFGEN models as input SEDs yields photometric temperatures with random 
uncertainties of 1\,000-2\,000 K and systematic differences of $\lesssim$~1\,000~K with respect to the spectroscopic 
temperatures (based on the Martins et al. 2005 calibration). Kurucz SEDs yield similar random uncertainties but a 
larger systematic difference.

\section{Extinction law}

	The above results for low extinction are of little practical value for Galactic stars because most O stars
in the Milky Way are moderately or heavily extinguished, with an extinction law that varies for different
sightlines. Accurately accounting for extinction is the last required step in the process and, arguably, the hardest
one to fulfill. The most commonly used family of optical/NIR Galactic extinction laws is that of
Cardelli et al. (1989). However, that paper was based on an analysis of a small (29) sample of stars with only
moderate extinction and used an unphysical seventh degree polynomial in $1/\lambda$ for the optical region. Hence,
it is not possible to deredden observed photometry with the required accuracy for our goals 
($\lesssim$~1\%) using the Cardelli et al. (1989) extinction law for $E(B-V) \gtrsim 0.5$. For that reason we
believe that the time has come for the calculation of an alternative sightline-dependent Galactic optical/NIR
extinction law. Some work has been recently done in the NIR (Moore et al. 2005, Nishiyama et al. 2006) but the
optical has not been revised yet. We plan to do so in the incoming year using the data in the Galactic O star catalog
as input.


\begin{thebibliography}
  \bibitem{Bohl06}     Bohlin, R. C.               2006,  astro-ph/0608715
  \bibitem{Cardetal89} Cardelli, J. A. et al.      1989,  ApJ   345,  245
  \bibitem{Coheetal03} Cohen, M. et al.            2003,  AJ    126, 1090
  \bibitem{HolbBerg06} Holberg, J. \& Bergeron, P. 2006,  AJ    132, 1221
  \bibitem{Hummetal88} Hummer, D. G. et al.        1988,  ApJ   328,  704
  \bibitem{LanzHube03} Lanz, T. \& Hubeny, I.      2003,  ApJS  146,  417
  \bibitem{LanzHube06} Lanz, T. \& Hubeny, I.      2006,  astro-ph/0611891
  \bibitem{Maiz04}     Ma\'{\i}z Apell\'aniz, J.   2004,  PASP  116,  859
  \bibitem{Maiz05}     Ma\'{\i}z Apell\'aniz, J.   2005,  PASP  117,  615
  \bibitem{Maiz06a}    Ma\'{\i}z Apell\'aniz, J.   2006a, AJ    131, 1184
  \bibitem{Maiz06b}    Ma\'{\i}z Apell\'aniz, J.   2006b, astro-ph/0609430
  \bibitem{Martetal05} Martins, F. et al.          2005,  A\&A  436, 1049
  \bibitem{MartPlez06} Martins, F. \& Plez, B.     2006,  A\&A  457,  637
  \bibitem{Mooretal05} Moore, T. J. T. et al..     2005,  MNRAS 359,  589
  \bibitem{Nishetal06} Nishiyama, S. et al.        2006,  ApJ   638,  839
\end{thebibliography}
\end{document}